# A Framework for Data-Based Turbulent Combustion Closure: *A Posteriori* Validation

Rishikesh Ranade and Tarek Echekki[1]

*Department of Mechanical and Aerospace Engineering, North Carolina State University, Raleigh, NC 27695-7910, USA*

**Abstract**

In this work, we demonstrate a framework for developing closure models in turbulent combustion using experimental multi-scalar measurements. The framework is based on the construction of conditional means and joint scalar PDFs from experimental data based on the parameterization of composition space using principal component analysis (PCA). The resulting principal components (PCs) act as both conditioning variables and transport variables. Their chemical source terms are constructed starting from instantaneous temperature and species measurements using a variant of the pairwise mixing stirred reactor (PMSR) approach. A multi-dimensional kernel density estimation (KDE) approach is used to construct the joint PDFs in PC space. Convolutions of these joint PDFs with conditional means are used to determine the unconditional means for the closure terms: the mean PCs chemical source terms and the density. These means are parameterized in terms of the mean PCs using artificial neural networks (ANN). The framework is demonstrated *a posteriori* using the data from the Sandia piloted turbulent jet flames D, E and F by performing RANS calculations. The radial profiles of mean and RMS of temperature and measured species mass fractions agree well with the experimental means for these flames.

---

[1] Corresponding Author. Address: Department of Mechanical and Aerospace Engineering, North Carolina State University, 911 Oval Drive, Campus Box 7910, Engineering Building III, Room 3252, Raleigh, NC 27695-7910, USA. Fax: +1 919 515 7968, E-mail address: techekk@ncsu.edu (T. Echekki).

1. **Introduction**

One of the primary challenges in turbulent combustion modeling is the prediction of the so-called turbulence-chemistry interactions (TCI). In the averaged or filtered forms of the Navier-Stokes equations and thermo-chemical scalars' equations, these interactions translate into unclosed terms for the non-linear contributions to these equations. Some of the common approaches for modeling TCI are based on the parameterization of the composition space in terms of a reduced set of variables (e.g. the mixture fraction and reaction progress variable) from which the probability density functions (PDF) and the full set of thermo-chemical scalars are determined (e.g. [1] and [2]). Accordingly, the solution of reacting flows involves the transport of these parameters' moments' unconditional means. Key challenges in developing these models include the selection of the conditioning variables, the prescription of an accurate PDF and the determination of the reactor model (e.g. steady flamelet) needed to construct conditional means for the thermo-chemical scalars.

Nonetheless, these moment-based models tend to be more computationally efficient than more advanced models that seek to directly evaluate unclosed terms in the governing equations, such as the PDF transport approach [3], the LEM-LES [4] and the ODT-LES [5] approaches. The use of low-dimensional stochastic models, such as the linear-eddy model (LEM) [6] and the one-dimensional turbulence (ODT) model [7] to construct PDFs and conditional means also provide an *a priori* tabulation approach for both conditional statistics and PDFs [8-12].

Recently, we have proposed a novel framework within the context of Reynolds-averaged Navier-Stokes (RANS) or large-eddy simulation (LES) to develop closure models using experimental measurements [13]. It is formulated around the same principle of parameterizing the composition space in terms of a reduced set of parameters using principal component analysis (PCA). The resulting principal components (PCs) serve as the conditioning variables for constructing conditional means of thermo-chemical scalars and the variables on which the statistical distributions, the PDFs, are constructed starting form multi-scalar measurements. Such measurements have provided essential data for improving turbulent combustion models [14-16]. These measurements include major species and temperature that can adequately represent the complexity of the composition space. Of course, the ultimate goal of the proposed framework is that such measurements can be extended to more practical combustion devices, with adequate optical access, to derive models around their design conditions.

Our approach [13] is constructed from experimental data and is designed to accommodate conditions where the traditional physics based conditioning parameters, such as mixture fractions or progress variables, may not be adequate. Developing the framework with partial data (since not all species that define the chemical mechanism are measured) poses additional challenges, which must be overcome. These challenges are related primarily to the calculation of conditional and unconditional means of the PCs' chemical source terms in their transport equations. The multi-scalar measurements do not provide this information. Therefore, the recovery of missing species is a primary task. In Ref. [13], we proposed a variant of the pairwise mixing stirred reactor (PMSR) [17-18], that mixes and reacts states that belong to the same region of the composition space. These states are initialized with the experimental measurements. An *a priori* validation for this approach using Sandia flames D, E and F [19] ODT simulation data [20-21] shows that this procedure can be adequate in reconstructing species reaction rates.

This study is an extension of this recent work [13] where we attempt to demonstrate the framework in *a posteriori* RANS simulations of the Sandia flames D, E and F. These simulations involve the transport of PCs along with flow equations. In the following sections, we describe the closure approach in more detail and present results from *a posteriori* studies of Sandia flames D, E and F in terms of mean radial of axial profiles of the experimentally measured scalars.

In Section 2, we present the preprocessing steps, model set up and simulation elements. In Section 3, we present results from *a priori* and *a posteriori* validations of the framework. Finally, conclusions and a discussion of future work are presented in Section 4.

## 2. Model Formulation

In this section, we describe the modeling framework's key elements, which involves the key preprocessing steps to build the model (PCA on the experimental data, the calculation of conditional means and the joint PDFs and the recovery of missing species) and the model set up and simulation. During the preprocessing step, *a priori* validation may be carried out as presented in Ref. [13]. This validation is an important step to evaluate the data adequacy in terms of size and ability to span the composition space and to determine the number of PCs that need to be retained from the PCA analysis.

### *2.1. PCA parameterization and governing equations*

For the closure model presented here, the RANS formulation is used. Although, a similar procedure can be implemented for LES based on multi-scalar line measurement data and the construction of filtered probability density functions. As described earlier, PCs are a representation of the thermo-chemical scalars in a lower dimensional space [13, 22-30]. The vectors of PCs and their chemical source terms are linearly related to the vectors of measured thermo-chemical scalars and their chemical source terms, respectively:

$$\boldsymbol{\phi} = \mathbf{A}^T \boldsymbol{\theta}, \tag{1}$$

$$\mathbf{s}_{\boldsymbol{\phi}} = \mathbf{A}^T \mathbf{s}_{\boldsymbol{\theta}} \tag{2}$$

where, $\boldsymbol{\theta} = (T, Y_1, Y_2, \ldots, Y_{N-1})$ is the vector of $N$ measured thermo-chemical scalars (temperature and $N-1$ species). $\boldsymbol{\phi} = (\phi_1, \phi_2, \ldots, \phi_{N_{PC}})$ is the vector of the $N_{PC}$ corresponding PCs. The measured scalars tend to include major species and temperature, which are representative of the composition space accessed by the measurements. The PCs are ordered based on the magnitude of their associated eigenvalues from highest to lowest, which also measures their relative contribution to the data variance. Here, the first $N_{PC}$ PCs out of $N$ carry an acceptable cumulative contribution to the data variance, which is typically 99% or higher for model implementation. The transformation matrix, $\mathbf{A}^T$, is made up of the first $N_{PC}$ eigenvectors of the $N \times N$ data's covariance matrix, which also corresponds to the retained PCs. $\mathbf{s}_{\boldsymbol{\theta}}$ and $\mathbf{s}_{\boldsymbol{\phi}}$ correspond to the chemical source terms for the thermo-chemical scalars and the PCS, respectively. They, too, are related through matrix $\mathbf{A}^T$ as indicated in Eq. (2) [22].

The governing equations include 1) continuity, 2) linear momentum and 3) the PCs transport equations are:

Continuity:
$$\frac{\partial \bar{\rho}}{\partial t} + \frac{\partial \bar{\rho} \tilde{u}_j}{\partial x_j} = 0 \tag{3}$$

Momentum: $$\frac{\partial \bar{\rho}\tilde{u}_i}{\partial t} + \frac{\partial \bar{\rho}\tilde{u}_i\tilde{u}_j}{\partial x_j} = -\frac{\partial \bar{p}}{\partial x_i} + \frac{\partial}{\partial x_j}\left[2\bar{\rho}(\nu_T + \nu)\widetilde{S}_{ij}\right], i = 1,2,3. \qquad (4)$$

PCs: $$\frac{\partial \bar{\rho}\tilde{\phi}_k}{\partial t} + \frac{\partial \bar{\rho}\tilde{u}_j\tilde{\phi}_k}{\partial x_j} = \frac{\partial}{\partial x_j}\left[\bar{\rho}\left(\frac{\nu_T}{\mathrm{Sc}_T} + \nu_k\right)\frac{\partial \tilde{\phi}_k}{\partial x_j}\right] + \bar{s}_{\phi_k}, k = 1,\cdots, N_{\mathrm{PC}} \qquad (5)$$

In the above expressions, the symbols " $\bar{\ }$ " and " $\sim$ " correspond to Reynolds averaging and density weighted averaging, respectively. $\tilde{u}_i$ and $\tilde{\phi}_k$ are the Favre-averaged velocity component in the $i$th direction and the $k$th PC, respectively. $\nu_T$ is a turbulent kinematic viscosity, $\mathrm{Sc}_T$ is a turbulent Schmidt number and $S_{ij}$ is the $ij$ component of the rate-of-strain tensor. Within the context of the presented governing equations, the turbulent kinematic viscosity, $\nu_T$, is obtained from turbulence closure (e.g. the $k-\epsilon$ model). Moreover, closure is required for the mean PC source terms, $\bar{s}_{\phi_k}$, the mean density, $\bar{\rho}$, to solve the governing equations. For low Mach number flows, the continuity equation is replaced by the Poisson equation and the density is recovered from the closure model. We also prescribe models for the Favre-averaged thermo-chemical variables, $\widetilde{\boldsymbol{\theta}}$, including temperature and experimentally measured species mass fractions in terms of the transported PCs. These models are needed to establish comparisons with the experimental data.

A PCA implemented on the experimental data enables the extraction of both the conditional means of the measured scalars and their derived functions (e.g. chemical source terms) and the joint PC PDFs. The unconditional statistics for the density, $\bar{\rho}$, the thermo-chemical scalars, $\tilde{\theta}_k$, and the PCs' source terms, $\bar{s}_{\phi_k}$, are determined through a convolution of their conditional means, $\langle\rho|\boldsymbol{\phi}\rangle$, $\langle\theta_k|\boldsymbol{\phi}\rangle$ and $\langle s_{\phi_k}|\boldsymbol{\phi}\rangle$, and the PDFs as follows:

$$\tilde{\theta}_k(\widetilde{\boldsymbol{\Phi}}) = \frac{\int \langle\rho|\boldsymbol{\phi}\rangle\langle\theta_k|\boldsymbol{\phi}\rangle\, P(\boldsymbol{\phi};\widetilde{\boldsymbol{\phi}})d\boldsymbol{\phi}}{\bar{\rho}}, \text{ where } \bar{\rho} = \int \langle\rho|\boldsymbol{\phi}\rangle P(\boldsymbol{\phi};\widetilde{\boldsymbol{\Phi}})d\boldsymbol{\phi} \qquad (6)$$

$$\bar{s}_{\phi_k}(\widetilde{\boldsymbol{\Phi}}) = \int \langle s_{\phi_k}|\boldsymbol{\phi}\rangle\, P(\boldsymbol{\phi};\widetilde{\boldsymbol{\Phi}})d\boldsymbol{\phi} \qquad (7)$$

$P(\boldsymbol{\phi};\widetilde{\boldsymbol{\Phi}})$ is the joint PDF. The conditional means for the chemical source terms, $\langle s_{\phi_k}|\boldsymbol{\phi}\rangle$ requires the recovery of the missing species needed in a chemical mechanism to compute the thermo-chemical scalars source terms. Note, that the expression for the PDF also can include unconditional means for variances or higher moments. However, in our present analysis based on *a priori* [13] and *a posteriori* validation (presented here), the unconditional means of the PCs, along with the instantaneous values of the PCS, are sufficient to tabulate the PDFs.

## 2.2. Preprocessing Procedure

Figure 1 summarizes the preprocessing steps for developing the closure elements of the proposed framework starting with multi-scalar measurements. The steps (also indicated in the figure) are implemented as follows:

1. (Step 1) Starting from instantaneous multi-scalar measurements, $\boldsymbol{\theta}$ at different positions in the flow, we carry out a PCA to on the entire dataset to determine the matrix $\mathbf{A^T}$ and the PCs, $\boldsymbol{\phi}$ (step 1.1). Based on their cumulative contribution to the data variance, a set of $N_{PC}$ PCs is retained (step 1.2). The adequacy of $N_{PC}$ can be evaluated through *a priori* analysis of the framework (see Ref. [13]). The next stage involves the determination of the conditional means of thermo-chemical scalars (step 1.3). Finally, the unconditional means for the measured thermo-chemical scalars and density (step 1.4) are evaluated. These unconditional means are "tabulated" vs. the PCs' unconditional means using artificial neural networks (ANN). As a multi-variate, non-linear regression method, ANN has found use in different applications in combustion [31-44].

   Although, potentially, we can compute these means using a convolution of the conditional means from step 1.3 and the PCs' PDF (from step 3 below) using Eq. (6) and similar to the approach shown in Fig. 1, we have chosen in this study to compute the thermochemical scalars' unconditional means directly by averaging the available single-shot data. For Favre averages, these averages are weighted by the mixture density. In this study, the evaluation of Eq. (6) by convolution is primarily used to validate the PDF tabulation. This PDF tabulation is essential to evaluate the PCs' chemical source terms carried out in step 2 using Eq. (7) since these terms are not readily available from the experimental data. Of course, the construction of unconditional means through a convolution of the conditional means provides a useful generalization of the modeling framework.

2. (Step 2) Next, we carry out a modified PMSR. The 'particles' correspond to states that are initialized from the experimental measurements. They are divided into clusters based on their proximity in composition space (step 2.1). The PMSR is implemented in each cluster (step 2.2). The PMSR calculations recover the missing species needed to determine the thermo-chemical scalars' chemical sources in chemical mechanism (step 2.3). The determination of these source terms is an essential step towards computing the PCs chemical source terms using Eq. (2). Upon the completion of the PMSR calculations in the different clusters, every reconstructed thermo-chemical scalar field can be associated with a vector of PCs and their chemical source terms. Conditional averages of the PCs chemical source terms, $\left\langle s_{\phi_k} \middle| \boldsymbol{\phi} \right\rangle$, are computed in terms of the retained PCs using the data evolved from the PMSR calculations (step 2.4). Again, the conditional means are constructed using the entire data without distinction of its spatial origin. Finally, the PCs' chemical source terms unconditional means, $\bar{s}_{\phi_k}(\widetilde{\boldsymbol{\phi}})$, in terms of the PCs' unconditional means are evaluated using the convolution of their conditional means and the PCs' PDFs based on Eq. (7) (step 2.5). Again, ANN regression is used to "tabulate" $\bar{s}_{\phi_k}(\widetilde{\boldsymbol{\phi}})$.

3. (Step 3) The instantaneous PCs at different axial and radial positions are used to construct their joint PDFs based on their corresponding multiple shots of data using the kernel density estimation (KDE) approach [45]. KDE offers a smooth reconstruction of the joint distribution. KDE is a non-parametric way to estimate the PDF of a random variable. A *d*-dimensional kernel density function (KDF) is expressed as a sum of kernel functions centered on model determined points, which are learned from sample data [45]:

$$P(\boldsymbol{\phi}; \widetilde{\boldsymbol{\phi}}) = \frac{1}{nh}\sum_{i=1}^{n} K\left(\frac{\boldsymbol{\phi}-\widehat{\boldsymbol{\phi}}_i}{h}\right) \quad (8)$$

where, $K$ is the kernel function, $h$ is the bandwidth and $\widehat{\boldsymbol{\phi}}_i$ are the $n$ samples of the selected PCs for a given condition, $\widetilde{\boldsymbol{\phi}}$. $\widetilde{\boldsymbol{\phi}}$ is the Favre average of a PC vector $\boldsymbol{\phi}$ at a given position in space; although, if the same averages are available at different positions, a single joint PDF is constructed for them. A common kernel function, which is used in this study, is the Gaussian distribution.

The bandwidth controls the amount of smoothing of the kernel function. Larger bandwidths lead to smoother functions that can capture the global trends while smaller bandwidths can capture the sharp changes observed in turbulence combustion data. We have observed that bandwidths ranging between 0.005 and 0.05 for a Gaussian kernel function are enough to smoothly capture the important characteristics present in the experimental data. The joint KDE is implemented in python using the in-built functions available in their *scikit-learn* library [46].

The use of KDE in combustion is relatively new. However, it was demonstrated by our group in two recent studies [12, 13] to construct marginal uni-variate PDFs and joint PDFs. The PDFs, of course, have to be determined using multiple measurements at one point. In principle, once these PDFs are constructed at the different measurement points, we can develop a general regression for the PDF from these KDE-based PDFs, which is a function of the instantaneous retained PCs and parameterized with the unconditional means for these PCs. However, in the results presented in Section 3.4, we use the PDFs from KDE at the different measurement positions to determine the unconditional means for the closure term in the governing equations: the density and the PCs source terms.

As discussed earlier, the experimental data does not provide any information related to the chemical source terms of the measured temperature and species. As proposed in [13], a PMSR model is used to reconstruct the missing species in the chemical mechanism used to predict the measured species chemical source terms. A PMSR [17,18] is a zero-dimensional stochastic reactor model in which a finite number of particles starting from an initial composition are evolved by reaction and mixing. The reaction is modeled deterministically by integration of the reaction source terms from a reaction mechanism while the mixing is carried out stochastically by pairing particles and mixing them over a specified 'mixing' time, $\tau_m$. In [13], we demonstrated the ability of the modified PMSR model to reconstruct missing species and determine the chemical source terms for the measured quantities in flames D, E and F using data from ODT simulations of the same flames [20-21]. As a low-dimensional model, ODT has been successfully implemented to study jet diffusions flames [47-52], including the Sandia flames [20-21]. The model can provide statistics, including conditional means and PDFs, as well as instantaneous data that can be used to determine thermo-chemical scalars' chemical source terms.

The main distinction of the PMSR approach [13] compared to the original PMSR formulation [17,18] is that all particles are simultaneously used in the reactor in our approach; while, in the original PMSR model, new particles are introduced into the reactor at prescribed intervals and an equal number is removed at the same time. Moreover, a unique implementation of the PMSR within the context of the present framework is the choice of clusters of particles within the same region of the composition space. In Ref. [13] the clusters are chosen strictly based on the values of their corresponding PCs. A simple binning procedure in PC-space can then determine automatically where every measurement or particles fits in the different clusters. This choice may be a natural one. However, it does not guarantee that adequate sizes for the clusters (i.e. the number of particles) are achieved based on the adequacy of the data. In this study, we propose an alternative clustering based on Kohonen self-organizing maps (SOM) [53]. This method has been successfully implemented in combustion applications ([36,54]). In the Kohonen SOM algorithm, the input vectors of the measured species and temperature are assigned a cluster based on the Euclidian distance. A node is assigned to each cluster and is representative of the thermo-chemical composition in it. The nodes are arranged as 2-D maps and the SOM creates topologically ordered mappings between the nodes and the input data.

## 3. Results and discussion

In this section, we present results of our closure methodology implemented on Sandia flames D, E and F.

### 3.1. Run conditions

The Sandia flames D, E and F [19] are well-characterized turbulent piloted jet partially premixed flames. The flames have similar inlet compositions and temperature but differ in flow conditions. The burner consists of a central fuel jet with an inner diameter of 7.2 mm which is surrounded by a concentric pilot jet with inner and outer dimeters of 7.7 and 18.2 mm respectively. The fuel jet is a mixture of 75% $CH_4$ and 25% air with a temperature of 294 K. On the other hand, the pilot is based on unstrained premixed flame solution of methane-air flame at an equivalence ratio of 0.88 and temperature of 1880 K. The co-flow consists of air with a temperature of 291 K and velocity of 0.9 m/s. The fuel jet Reynolds numbers based on the jet diameter are 22,400, 33,600 and 44,800 for the flames D, E and F, respectively. The pilot jet velocities are 11.4, 17.1 and 22.8 m/s respectively. The experimental data corresponding to flames D, E and F consists of instantaneous measurements collected at different downstream and radial positions. Each measurement position includes from 300-1000 single shot multi-scalar measurements.

### 3.2. RANS solution

In this work, a realizable $k$-$\varepsilon$ model it used to formulate the turbulent viscosity [55] because of its proven accuracy in predicting round and planar jets. Hence, additional transport equations for the turbulent kinetic energy, $q$, and its dissipation rate, $\epsilon$, are needed in addition to the transport equations for mass, momentum and the PCs (Eqs. (3) – (5)) as prescribed below:

$q$:
$$\frac{\partial \bar{\rho}q}{\partial t} + \frac{\partial \bar{\rho}q\tilde{u}_j}{\partial x_j} = \frac{\partial}{\partial x_j}\left[2\bar{\rho}(\nu_T + \nu)\frac{\partial q}{\partial x_j}\right] + G_k + G_b - \bar{\rho}\tilde{\epsilon} \tag{9}$$

$\epsilon$:
$$\frac{\partial \bar{\rho}\epsilon}{\partial t} + \frac{\partial \bar{\rho}\epsilon\tilde{u}_j}{\partial x_j} = \frac{\partial}{\partial x_j}\left[2\bar{\rho}(\nu_T + \nu)\frac{\partial \epsilon}{\partial x_j}\right] + \bar{\rho}C_1 S\epsilon - \bar{\rho}C_2\frac{\tilde{\epsilon}^2}{q+\sqrt{\nu\epsilon}} + C_{1\epsilon}\frac{\epsilon}{q}C_{3\epsilon}G_b \tag{10}$$

The solutions for $q$ and $\epsilon$ are used to determine the turbulent kinematic viscosity: $\nu_T = C_\mu \frac{q^2}{\varepsilon}$, where $C_\mu$ is not a constant as in the standard $k$-$\varepsilon$ approach [55]. Also, $S$ is expressed in terms of the inner product of $S_{ij}$, $S = \sqrt{2S_{ij}S_{ij}}$ and $C_2, C_{1\epsilon}$ $\sigma_k$ and $\sigma_\epsilon$ are model constants set to 1.9, 1.44, 1.0 and 1.2 respectively. The details of all the other model definitions and constants can be found in Ref. [55]. The choice of the turbulence model is dictated primarily by the model's ability to describe the growth of the jet and to capture the downstream evolution of the mixture fraction mean profiles.

The RANS simulation is carried out in Ansys Fluent 19.0. In the Fluent solver, the governing equations are solved in their conservative form, as described previously in Eq. (3) to Eq. (5), Eq. (8) and Eq. (9). The simulation was set-up on a 2-D axis-symmetric domain and a structured, quadrilateral mesh with a cell count of close to 150,000 cells. A highly refined mesh is considered to ensure a grid independent solution. The closure model is integrated into the solver by using user-defined functions (UDFs). The pressure-velocity coupling was modeled using a SIMPLE scheme. The solution for the dynamic pressure is designed to be a solution for the continuity equation, since density is provided through the PC solution as prescribed in Eq. (6). First order upwind schemes were used for $q$, $\epsilon$ and the PCs transport equations; while the pressure and momentum equations are resolved to second order accuracy. The solution convergence is tracked by monitoring the residuals of all transported quantities. The solution converges when the residual falls below $10^{-5}$. Since this is a steady state run, an initial flame solution is provided. This solution is obtained by interpolating the radial profiles' solutions at different downstream distances.

Given that experimental data is available starting at $x/d$ = 7.5, we have chosen to start our simulations at this downstream distance instead of the inlet of the burner to enable a full data-based modeling approach. However, we still need to provide the inlet flow and turbulence conditions at this downstream distance, which is not provided by the experimental data. These conditions are evaluated by carrying out a flamelet-generated model (FGM) simulation of the inlet conditions from $x/d$ = 0 to $x/d$ = 7.5 (https://www.fgm-combustion.org/downloads) [56] using the recommended values for the inlet conditions at the Turbulent Nonpremixed Flames (TNF) workshop. The solutions of the flow and turbulence ($\tilde{u}_i$, $q$ and $\epsilon$) at $x/d$ = 7.5 from FGM RANS simulations is used for the inlet of the present simulation. The turbulent Schmidt numbers for the PCs, $\phi_1, \phi_2$ and $\phi_3$ are set to 0.05, 0.1 and 0.05 respectively.

### 3.3. Some results of the preprocessing steps

PCA is carried out on all flames at all measurement positions. The measured quantities are treated as the representative scalars of the composition space. The physical significance of PCs for the three flames considered here has already been discussed in Ref. [13], indicating the relevance of the first PCs to a reaction progress variable and mixture fraction, respectively; while, the third PC includes measures of chemical reactivity through the intermediates. In the present study, we retain 3 PCs, which account for approximately 99% of the data variance. The range of the 3 PCs, $\phi_1, \phi_2$ and $\phi_3$ is divided into 20 bins each and conditional means are evaluated within these bins.

In this work, the ANN training for tabulation of the conditional and unconditional means is carried out using the Bayesian-Regularization training algorithm in Matlab 2017. Separate networks are constructed for different output variables. The PC source terms networks consist of input and output layers and 3 hidden layers; and the number of neurons in each hidden layer is 30, 22 and 15, respectively. The thermo-chemical scalar networks are simpler and contain, again, an input and an output layer and 2 hidden layers with 30 and 15 neurons, respectively. The number of hidden layers and neurons is selected based on a 4-fold cross validation approach. The data set is divided into training, validation and testing set in the following ratio, 70/15/15. The ANN training is allowed to run until the mean squared error between the network output and target value for the testing data dropped below $10^{-6}$ or the validation error continues to increase for 6 consecutive iterations. The entire training process takes approximately an hour on a single processor of an Intel Xeon CPU.

### 3.3.1. Data Clustering with SOM

For implementation of the missing species recovery in the modified PMSR, the instantaneous PCs are grouped into 64 cluster nodes on an $8 \times 8$ 2D lattice, with each node containing from 300-3000 particles. This choice is made to keep a reasonable number of particles within each cluster. Other factors may play a role in determining the number of clusters, including 1) the localness within the composition space for the PMSR particles and 2) the management of the computational cost associated with PMSR simulations. Figure 2 shows the distribution of the PCs in each cluster among the 2D lattice. It is important to note that the transition between neighboring clusters is very smooth for all 3 PCs. The chemistry integration is implemented using the reduced 12-step augmented reduced mechanism (ARM) for methane oxidation [57].

Figure 2 shows the average values of the PCs within each cluster on the 2D lattice. The figure clearly shows how smooth transitions occur between contiguous nodes demonstrating how well SOM is partitioning the composition space.

### 3.3.2. Conditional Means

The conditional means for temperature, CO and OH mass fractions are shown in Figs. 3-5, respectively, on 2D contours for, $\phi_1$ and $\phi_2$ and different values for the third PC, $\phi_3$. From Ref. [13], we have determined that $\phi_1$, $\phi_2$ and $\phi_3$ are aligned with a reaction progress variable, a mixture fraction and intermediates or a measure of reactivity, respectively. From Figs. 4 and 5, we can see that CO and OH exhibit a strong dependence on $\phi_3$ given the strong correlation of this PC with these species. In contrast, the temperature (and reactants and products not shown here) exhibits similar contours at different levels of $\phi_3$. Nonetheless, for both sets of statistics, the accessed composition space varies with these levels.

This composition space can be augmented with supplemental data if needed, including conditions for pure mixing outside the reaction zone and equilibrium downstream of the flame. However, neither strategy is adopted here.

*3.3.3.. PC source terms*

The conditional means of PC source terms are shown in Figs. 6-8 on 2D contours for $\phi_1$ and $\phi_2$ and different levels for the third PC, $\phi_3$. It may be observed that all the source terms have a strong dependence on $\phi_3$. This trend is expected since we have identified this PC as a measure of chemical reactivity. It is correlated with intermediates that peak in the reaction zone. Hence the 3$^{rd}$ PC needs to be retained for *a posteriori* analysis. In contrast, we have shown in Ref. [13] that 2 PCs are adequate to reproduce other statistics, such as PDFs and unconditional means for the measured scalars. Also, and perhaps this is also expected, the chemical source terms for the 3 leading PCs tend to exhibit peaks and minima within the same regions of the composition space, which may be aligned with different layers of the reaction zone.

*3.4. A posteriori results*

Finally, we present *a posteriori* results using the RANS model described in Eqs. (4), (5), (9) and (10). The mean radial profiles of the temperature, *T*, the mixture fraction, *Z*, and the CO, $H_2$, OH, $H_2O$ and $CO_2$ mass fractions are reported and compared against the experimental means for all flames D, E and F. These comparisons are made at axial distances of *x/d* = 15, 30, 45 and 60 and shown in Figs. 9-11. The mixture fraction is based on the Bilger's definition for the mixture fraction [58]. It is included to identify the governing equations adequacy in predicting the mixing and the jet growth process. These predictions are an essential step before we can establish comparisons of the radial profiles of the reactive scalars.

Overall, the radial profiles' comparisons of these reactive scalars between RANS simulations and experiments match reasonably well. However, the most prominent deviations from the experimental data occur either at *x/d* = 15 or *x/d* = 60 for CO, $H_2$ and OH. In contrast, the temperature and the products mass fractions are in excellent agreement across all downstream conditions. However, in all cases, a comparison between flames D, E and F show the clear trends as the jet Reynolds number is increased, including 1) lower values for the intermediates at *x/d* = 15, a condition of extinction, and 2) slightly broader radial profiles for all scalars at all shown downstream distances.

Also, at this downstream distance, the CO, OH and $H_2$ predictions for flames E and F are in better agreement with the experimental data compared to flame D. This may appear to be counterintuitive as many models tend to predict flame D results better than the higher Reynolds number flames; but, it is important to emphasize that the model is based on data that is collected across different flame conditions.

Uncertainties in the predictions of the present data-based framework can be associated with different factors, including 1) the experimental data uncertainty, which is typically within a few percents, and adequacy, 2) the number of PCs retained, and 3) the closure terms, including primarily the closure for the PCs chemical source terms. The adequacy of the data is closely related to the model reduction, which is associated with the number of PCs retained. A simple scaling argument suggests that the required size of the data scales with the required data per PC to a power that corresponds to the number of these retained PCs. Although, one must consider the data variability in any given dimension and for any given PC. In the *a priori* study presented in Ref. [13], we have demonstrated a clear distinction in the predictions based on 2 vs. 4 PCs, with the fourth PC closely correlated with OH and non-negligible contributions from CO and $H_2$. The trade-off between the choice of 3 PCs for PC transport vs. 4 PCs is dictated, in addition to their ability to represent the composition space, by the size of the data (e.g. how well can we capture a multi-dimensional KDE with a given database size?).

The last mechanism for potential discrepancy is associated with the prediction of the PCs source terms. The quality of this prediction depends on the experimental uncertainty and the adequacy of the PMSR model in the recovery of the missing species and in the evaluation of the thermo-chemical scalars chemical source terms and subsequently the PCs chemical source terms.

Finally, the RMS profiles of the thermo-chemical scalars can be determined using similar regressions for the RMS values vs. the unconditional means for the PCs. These profiles are shown in Figs. 12-14 for flames D, E and F for the temperature and measured species mass fractions. The same conclusions associated with the results of Figs. 9-11 can be extended for the RMS profiles showing were excellent agreements with experimental statistics are found and where there are discrepancies. Nonetheless, the overall performance of the experiment-based modeling framework yields satisfactory results.

## 4. Conclusion and Future Extensions

In this study, we have demonstrated, using *a posteriori* simulation, a framework to construct turbulent combustion models starting with multi-scalar experimental measurements. The closure model is based on the parameterization of the composition space using PCs from which conditional means of the measured scalars and joint PCs PDFs, using KDE, can be constructed from experimental data. We also have implemented an approach, based on a variation on the PMSR model, to recover missing species in measurements, which are needed to calculate the PCs conditional and unconditional averages.

The closure model is validated using RANS simulations of the piloted jet Sandia flames D, E and F. The mean radial profiles for temperature, mixture fraction and other measured species agree reasonably well with the corresponding experimental statistics and capture the variation of these statistics from flames D, E and F associated with the presence of extinction and reignition. A similar agreement is found based on the comparisons of RMS of temperature and species under the same conditions. The RMS profiles are obtained, similarly to the mean profiles, using regressions vs. the unconditional means for the PCs. A trivial extension of the proposed framework, which will be implemented next, is its implementation within the context of LES. Here, line multi-scalar measurements can provide a ready source of data to construct filtered probability density functions (FPDFs).

There are additional framework extensions that also need to be addressed. The framework depends largely on the adequacy and accuracy of the experimental data. In order to develop conditional means and joint PDFs, a sufficient amount of data must be available. Of course, such data can be augmented with synthetic data to include conditions of pure mixing as well as equilibrium for bounding the composition space. There is equally extensive literature on the generation of synthetic data that is consistent with the prevailing statistics. However, one should attempt to do this only when other safer strategies are exhausted. For example, an effective evaluation of the joint PCs PDF can be carried out starting with individual PCs marginal PDFs. Such a strategy would be trivial if the PCs are statistically independent, which may or may not be the case. However, these PCs can be "rotated" using independent component analysis (ICA) to establish independent components (ICs) [59]. The treatment of the ICs is similar to that of PCs as far as determining their source terms given the linear relation between ICs and the measured scalars. Alternatively, the technique of Copula also establishes a relation between a joint PDF and marginal PDFs in multivariate systems [60]. For conditional means, a PMSR-like approach can mix different states in PC-space to fill the accessed composition space. Regardless, a close collaboration with the experimentalists is needed, especially when 3 or more PCs are needed to capture the data complexity and advance the solution in RANS or LES.

As stated above, this is a first step towards establishing a framework for experiment-based turbulent combustion modeling. Additional validation and extensions may be needed to accommodate additional modeling requirements, such as the framework's implementation for LES, for complex fuels and for more complex and practical combustor configurations.

**References**

1. N. Peters, Laminar diffusion flamelet models in non-premixed turbulent combustion, Prog. Energy Combust. Sci. 10 (1984) 319-339.
2. R.W. Bilger, Conditional moment closure for turbulent reacting flow, Phys. Fluids A: Fluid Dyn. 5 (1993) 436-444.
3. S.B. Pope, PDF methods for turbulent reactive flows, Prog. Energy Combust. Sci. 11 (1985) 119-192.
4. S. Menon, A.R. Kerstein, The linear-eddy model, in Turbulent Combustion Modeling (T. Echekki, E. Mastorakos, Eds.), pp. 221-247, Springer, Dordrecht (2011).
5. T. Echekki, A.R. Kerstein, J.C. Sutherland, The one-dimensional turbulence model in Turbulent Combustion Modeling (T. Echekki, E. Mastorakos, Eds.), pp. 249-276, Springer, Dordrecht (2011).
6. A.R. Kerstein, Linear-eddy model of turbulent scalar transport and mixing, Combust. Sci. Tech. 60 (1988) 391-421.
7. A.R. Kerstein, One-dimensional turbulence: Model formulation and application to homogeneous turbulence, shear flows, and buoyant stratified flows, J. Fluid Mech. 392 (1999) 277-334.
8. G. M. Goldin, A priori investigation of the constructed PDF model. Proc, Combust. Inst. 30(1) (2005) 785-792.
9. G.M. Goldin, S. Menon, Comparison of scalar PDF turbulent combustion models, Combust. Flame 113 (1998) 442-253.
10. V. Sankaran, T.G. Drozda, J.C. Oefelein, A tabulated closure for turbulent non-premixed combustion based on the linear eddy model, Proc. Combust. Inst. 32 (2009) 1571-1578.
11. W.H. Calhoon, S.F. Mattick, K. Kemenov, S. Menon, Strain effects in partially premixed methane-air jet flames. 53th Aerospace Sciences Meeting, Kissimmee, Florida, 5-9 January, 2015, AIAA Paper 2015-0674.
12. J.S. Miles, T. Echekki, A one-dimensional turbulence-based closure model for combustion LES, Combust. Sci. Tech. (2019) (https://doi.org/10.1080/00102202.2018.1556262).
13. R. Ranade, T. Echekki, A framework for data-based turbulent combustion closure: A priori validation, Combust. Flame. 206 (2019) 490-505.
14. R.S. Barlow, Laser diagnostics and their interplay with computations to understand turbulent combustion, Proc. Combust. Inst. 31 (2007) 1087-1095.
15. A.R. Masri, Design of experiments for gaining insights and validating modeling of turbulent combustion, in Turbulent Combustion Modeling: Advances, New Trends and Perspectives (Eds. T. Echekki and E. Mastorakos), Springer, Dordrecht, Netherlands, 2011.

16. S. Hochgreb, Mind the gap: Turbulent combustion model validation and future needs, Proc. Combust. Inst. 37:2091-2107 (2019).

17. S.B. Pope, Computationally efficient implementation of combustion chemistry using in-situ tabulation, Combust. Theo. Model 1 (1997) 41-63.

18. B. Yang, S.B. Pope, An investigation of the accuracy of manifold methods and splitting schemes in the computational implementation of combustion chemistry, Combust. Flame 112 (1998) 16-32.

19. Barlow, R.S., Frank, J.H., Effects of turbulence on species mass fractions in methane/air jet flames, Proc. Combust. Inst. 27 (1998) 1087-1095.

20. B. Ranganath, T. Echekki, One-dimensional turbulence-based closure with extinction and reignition, Combust. Flame 154 (2008) 23-46.

21. B. Ranganath, T. Echekki, ODT closure with extinction and reignition in piloted methane-air jet diffusion flames. Combust. Sci. Tech. 181(4) (2009) 570-596.

22. J.C. Sutherland, A. Parente, Combustion modeling using principal component analysis, Proc. Combust. Inst. 32 (2009) 1563-1570.

23. A. Parente, J.C. Sutherland, B.B. Dally, Tognotti, L., P.J. Smith. Investigation of the MILD combustion regime via principal component analysis, Proc. Combust. Inst. 33 (2011) 3333-3341.

24. A. Coussement, O. Gicquel, A. Parente, Kernel density weighted principal component analysis of combustion processes, Combust. Flame 159 (2012) 2844-2855.

25. H. Mirgolbabaei, T. Echekki, A novel principal component analysis-based acceleration scheme for LES–ODT: An a priori study, Combust. Flame 161 (2013) 898-908.

26. H. Mirgolbabaei, T. Echekki, Nonlinear reduction of combustion composition space with kernel principal component analysis, Combust. Flame 161 (2014) 118-126.

27. H. Mirgolbabaei, T. Echekki, N. Smaoui, A nonlinear principal component analysis approach for turbulent combustion composition space, Int. J. Hydrogen Energy 39 (2014) 4622-4633.

28. H. Mirgolbabaei, T. Echekki, The reconstruction of thermo-chemical scalars in combustion from a reduced set of their principal components, Combust. Flame 162 (2015) 1650-1652.

29. T. Echekki, H. Mirgolbabaei, Principal component transport in turbulent combustion: *a posteriori* analysis, Combust. Flame 162 (2015) 1919-1933.

30. O. Owoyele, T. Echekki, Toward computationally efficient combustion DNS with complex fuels via principal component transport, Combust. Th. Mod. 21(4) (2017) 770-798.

31. F. C. Christo, A. R. Masri, E. M. Nebot, T. Turanyi, Utilising artificial neural network and repro-modelling in turbulent combustion. In Neural Networks, Proceedings., IEEE International Conference on (Vol. 2, pp. 911-916) (1995) IEEE.
32. F. C. Christo, A. R. Masri, E. M. Nebot, S. B. Pope, An integrated PDF/neural network approach for simulating turbulent reacting systems, Symposium (International) on Combustion (Vol. 26, No. 1, pp. 43-48) (1996) Elsevier
33. F. C. Christo, A. R. Masri, E. M. Nebot, Artificial neural network implementation of chemistry with PDF simulation of $H_2/CO_2$ flames, Combust. Flame 106(4) (1996) 406-427.
34. J. A. Blasco, N. Fueyo, C. Dopazo, J. Ballester, Modelling the temporal evolution of a reduced combustion chemical system with an artificial neural network, Combust. Flame 113(1-2) (1998) 38-52.
35. J. A. Blasco, N. Fueyo, J. C. Larroya, C. Dopazo, Y. J. Chen, A single-step time-integrator of a methane–air chemical system using artificial neural networks, Computers & Chemical Engineering, 23(9) (1999) 1127-1133.
36. J. Y. Chen, J. A. Blasco, N. Fueyo, C. Dopazo, An economical strategy for storage of chemical kinetics: Fitting in situ adaptive tabulation with artificial neural networks, Proc. Combust. Inst. 28(1) (2000) 115-121.
37. J. A. Blasco, N. Fueyo, C. Dopazo, J. Y. Chen, A self-organizing-map approach to chemistry representation in combustion applications, Combust. Th. Mod. 4(1) (2000) 61-76.
38. M. Ihme, A. L. Marsden, H. Pitsch, H, Generation of optimal artificial neural networks using a pattern search algorithm: Application to approximation of chemical systems, Neural Computation 20(2) (2008) 573-601.
39. M. Ihme, C. Schmitt, H. Pitsch, Optimal artificial neural networks and tabulation methods for chemistry representation in LES of a bluff-body swirl-stabilized flame, Proc. Combust. Inst. 32(1) (2009) 1527-1535.
40. B. A. Sen, S. Menon, Linear eddy mixing based tabulation and artificial neural networks for large eddy simulations of turbulent flames, Combust. Flame 157(1) (2010) 62-74.
41. B. A. Sen, E. R. Hawkes, S. Menon, Large eddy simulation of extinction and reignition with artificial neural networks based chemical kinetics, Combust. Flame 157(3) (2010) 566-578.
42. A. K. Chatzopoulos, S. Rigopoulos, A chemistry tabulation approach via rate-controlled constrained equilibrium (RCCE) and artificial neural networks (ANNs), with application to turbulent non-premixed $CH_4/H_2/N_2$ flames, Proc. Combust. Inst. 34(1) (2013) 1465-1473.

43. R. Ranade, S. Alqahtani, A. Farooq, T. Echekki, An ANN based hybrid chemistry framework for complex fuels, Fuel 241 (2019) 625-636.
44. R. Ranade, S. Alqahtani, A. Farooq, T. Echekki, An extended hybrid chemistry framework for complex hydrocarbon fuels, Fuel 251 (2019) 276-284.
45. A.W. Bowman, A. Azzalini, Applied smoothing techniques for data analysis: The kernel approach with S-Plus illustrations, Oxford University Press, Oxford, 1997.
46. F. Pedregosa, G. Varoquaux, A. Gramfort, V. Michel, B, Thirion, O. Grisel, M. Blondel, P. Prettenhofer, R. Weiss, V. Dubourg, J. Vanderplas, A. Passos, D. Cournapeau, M. Brucher, M. Perrot, É. Duchesnay, Scikit-learn: Machine learning in Python, J. Mach. Learn. Res. 12 (2011) 2825−2830.
47. T. Echekki, A.R. Kerstein, T.D. Dreeben, J.-Y. Chen, 'One-dimensional turbulence' simulation of turbulent jet diffusion flames: Model formulation and illustrative applications, Combust. Flame 125 (2001) 1083-1105.
48. S. Zhang, T. Echekki, Stochastic modeling of finite-rate chemistry effects in hydrogen-air turbulent jet diffusion flames with helium dilution, Int. J. Hydrogen Energy 33 (2008) 7295-7306.
49. T. Echekki, K.G. Gupta, Hydrogen autoignition in a turbulent jet with preheated co-flow air, Int. J. Hydrogen Energy 34 (2009) 8352-8377.
50. K.G. Gupta, T. Echekki, One-dimensional turbulence model simulations of autoignition of hydrogen/carbon monoxide fuel mixtures in a turbulent jet, Combust. Flame 158 (2011) 327-344.
51. T. Echekki, S.F. Ahmed, Autoignition of n-heptane in a turbulent co-flowing jet, Combust. Flame 162 (2015) 3829-2846.
52. T. Echekki, S.F. Ahmed, Turbulence effects on the autoignition of DME in a turbulent co-flowing jet, Combust. Flame 178 (2017) 70-81.
53. T. Kohonen, J. Hynninen, J. Kangas, J. Laaksonen, SOM Pak: The self-organizing map program package. Report A31, Helsinki University of Technology, Laboratory of Computer and Information Science (1996).
54. L. L. Franke, A. K. Chatzopoulos, S. Rigopoulos, Tabulation of combustion chemistry via Artificial Neural Networks (ANNs): Methodology and application to LES-PDF simulation of Sydney flame L, Combust. Flame 185 (2017) 245-260.
55. T.H. Shih, W.W. Liou, A. Shabbir, Z. Yang, J. Zhu, A new $k$-$\varepsilon$ viscosity model for high Reynolds number turbulent flows, Comput. Fluids 24 (1995) 227-238.
56. W.J.S. Ramaekers, J.A. van Oijen, L.P.H. de Goey, A priori testing of flamelet generated manifolds for turbulent partially premixed methane/air flames, Flow. Turbul. Combust. 84 (2010) 439-458.

57. C.J. Sung, C.K. Law, J.-Y. Chen, Further validation of an augmented reduced mechanism for methane oxidation: Comparison of global parameters and detailed structure. Combust. Sci. Tech. 156 (2000) 201-220.
58. R.W. Bilger, S. Starner, R.J. Kee, On reduced mechanisms for methane-air combustion in nonpremixed flames, Combust. Flame 80 (1990) 135-149.
59. P. Comon, Independent component analysis—a new concept? Signal Proc. 36 (1994) 287–314.
60. C. Genest, A.-C. Favre, Everything you always wanted to know about Copula modeling but were afraid to ask, J. Hydro. Eng. 12 (2007) 347-368.

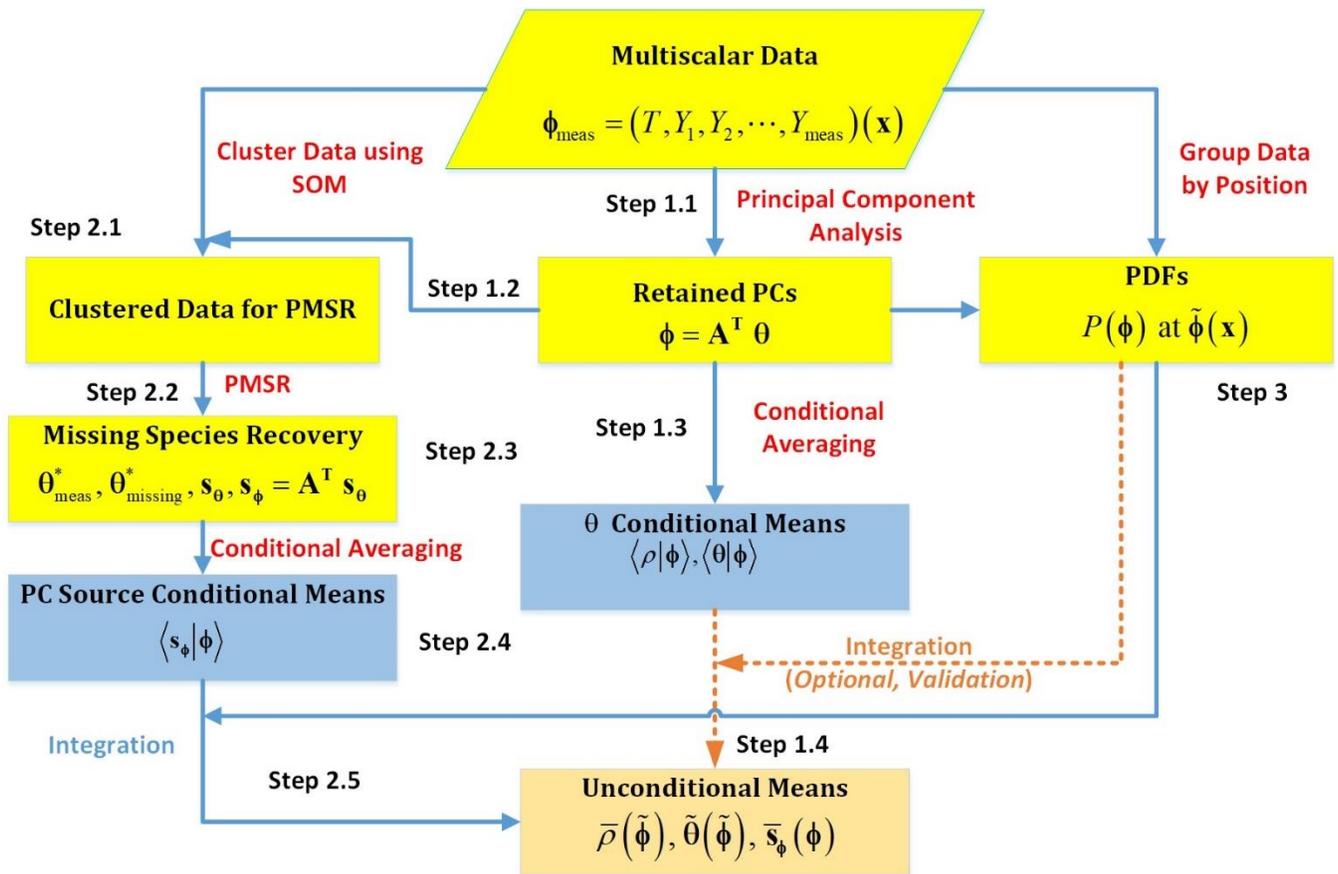

**Figure 1.** Flow chart of closure approach.

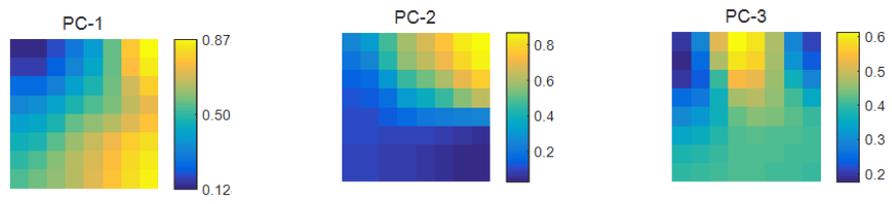

**Figure 2.** PC-distribution on the 2D SOM lattice.

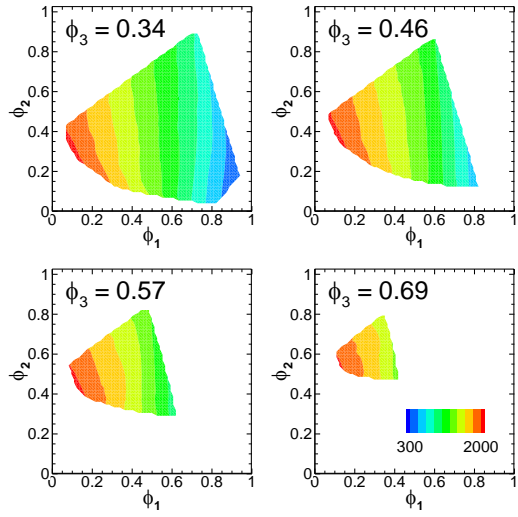

**Figure 3.** Conditional statistics iso-contours for temperature at different $\phi_3$ values.

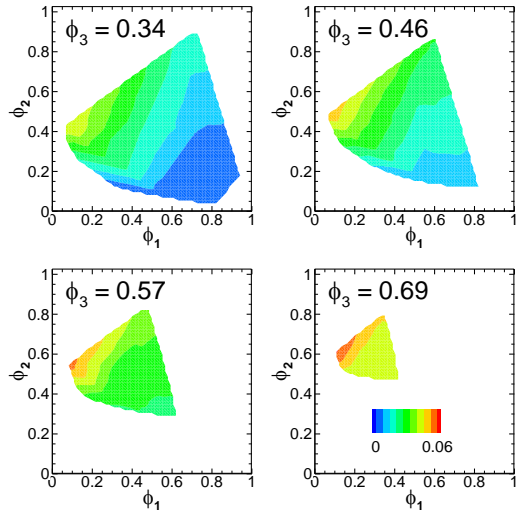

**Figure 4.** Conditional statistics iso-contours for CO mass fraction at different $\phi_3$ values.

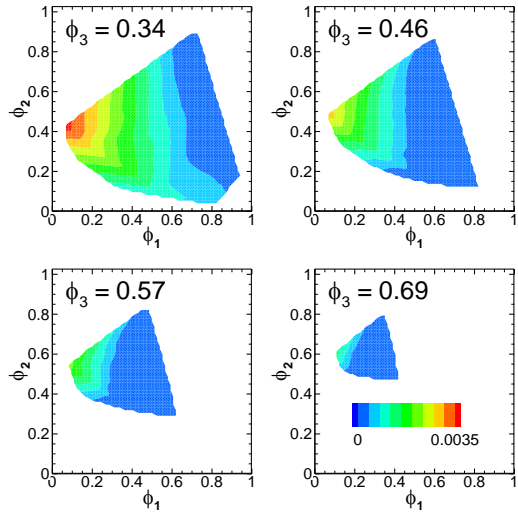

**Figure 5.** Conditional statistics iso-contours for OH mass fraction at different $\phi_3$ values.

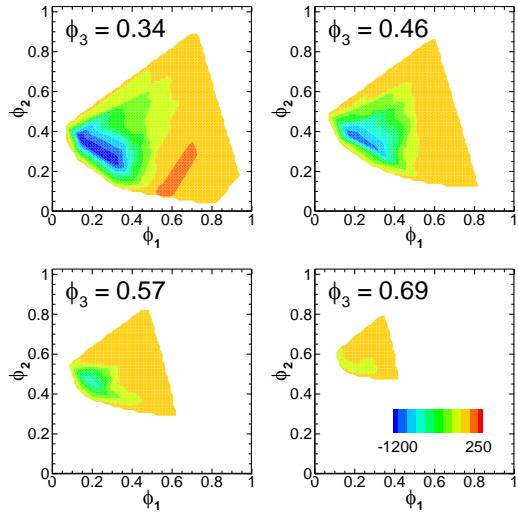

**Figure 6.** Conditional statistics iso-contours for $\phi_1$ source at different $\phi_3$ values.

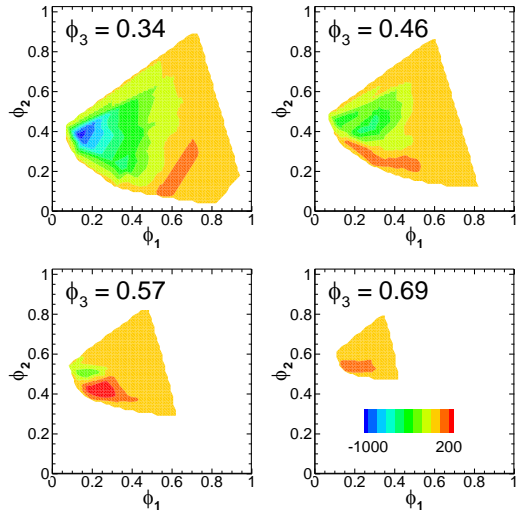

**Figure 7.** Conditional statistics iso-contours for $\phi_2$ source at different $\phi_3$ values.

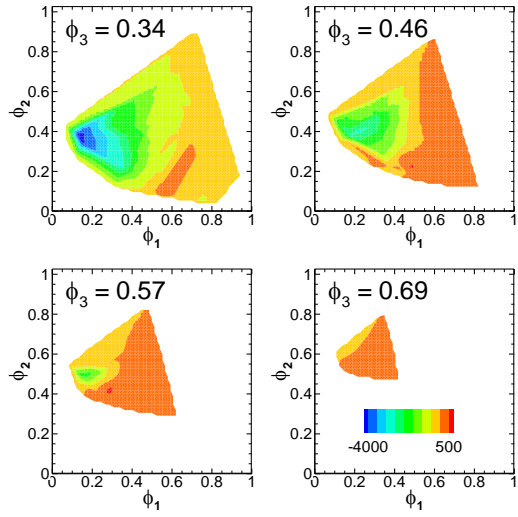

**Figure 8.** Conditional statistics iso-contours for $\phi_3$ source at different $\phi_3$ values.

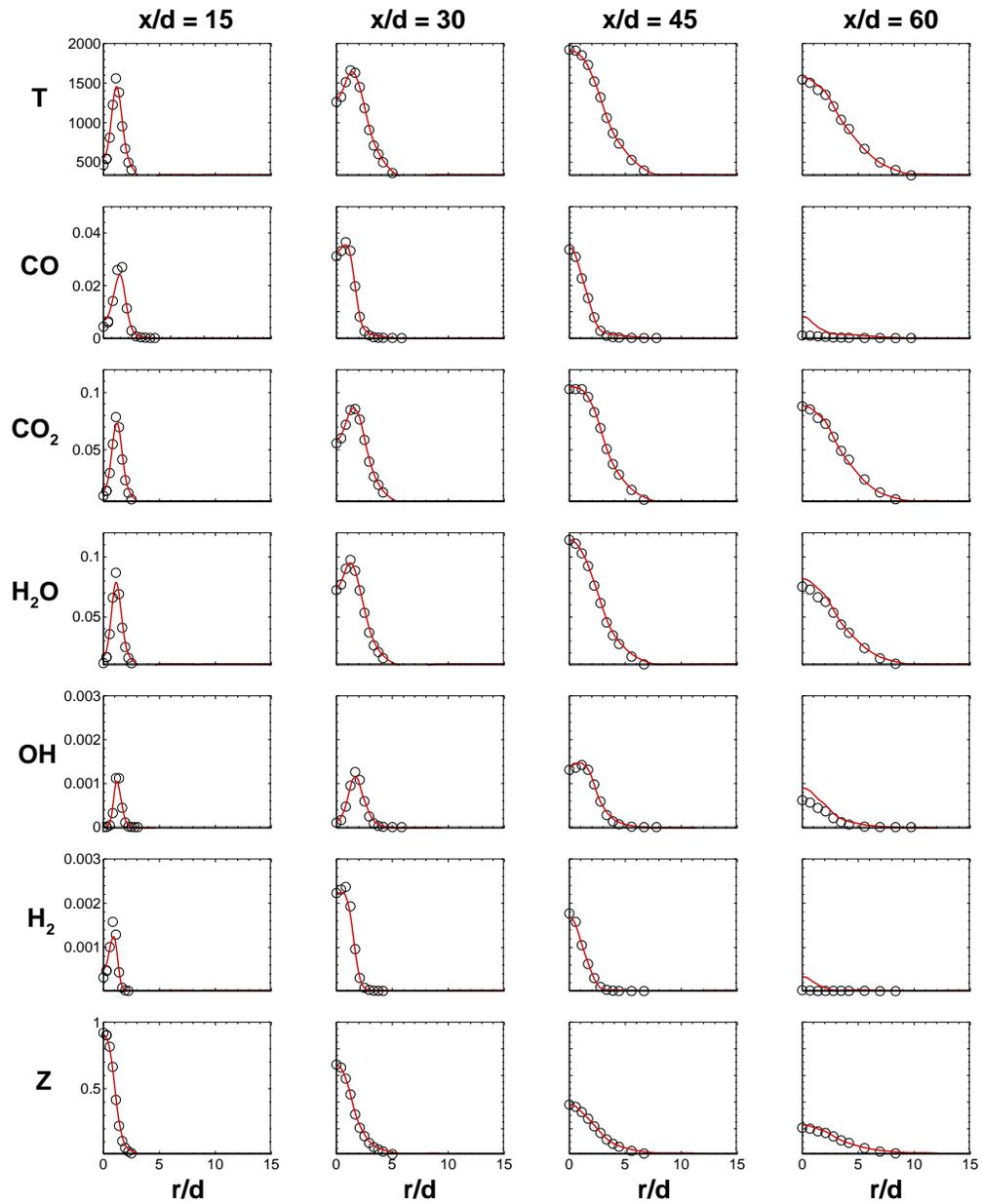

**Figure 9.** Flame D radial profile comparisons of experimental data (symbols) and the closure model (Red solid lines) of the Favre means of temperature, mixture fraction and measured species mass fractions at $x/d$ = 15, 30, 45 and 60.

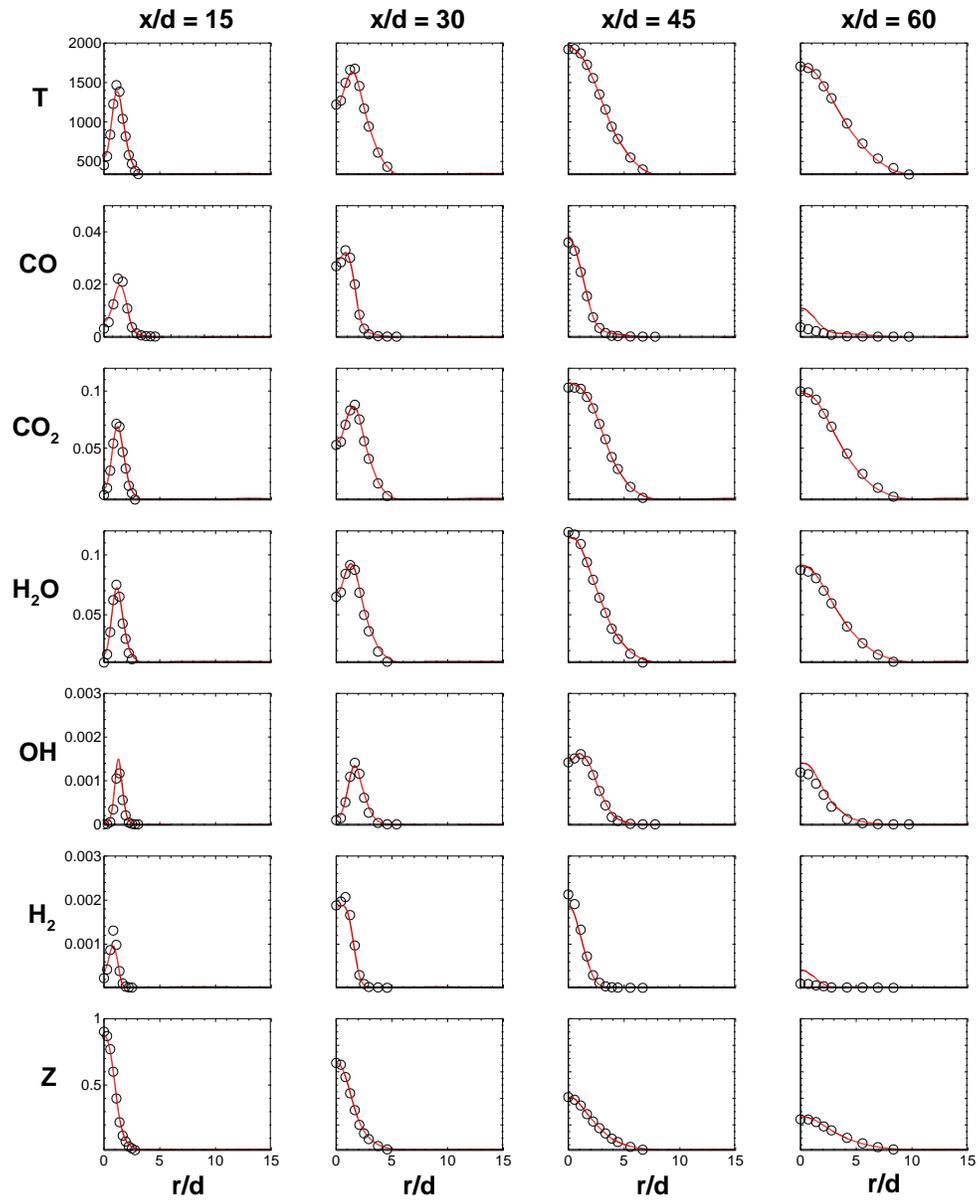

**Figure 10.** Flame E radial profile comparisons of experimental data (symbols) and the closure model (Red solid lines) of the Favre means of temperature, mixture fraction and measured species mass fractions at *x/d* = 15, 30, 45 and 60.

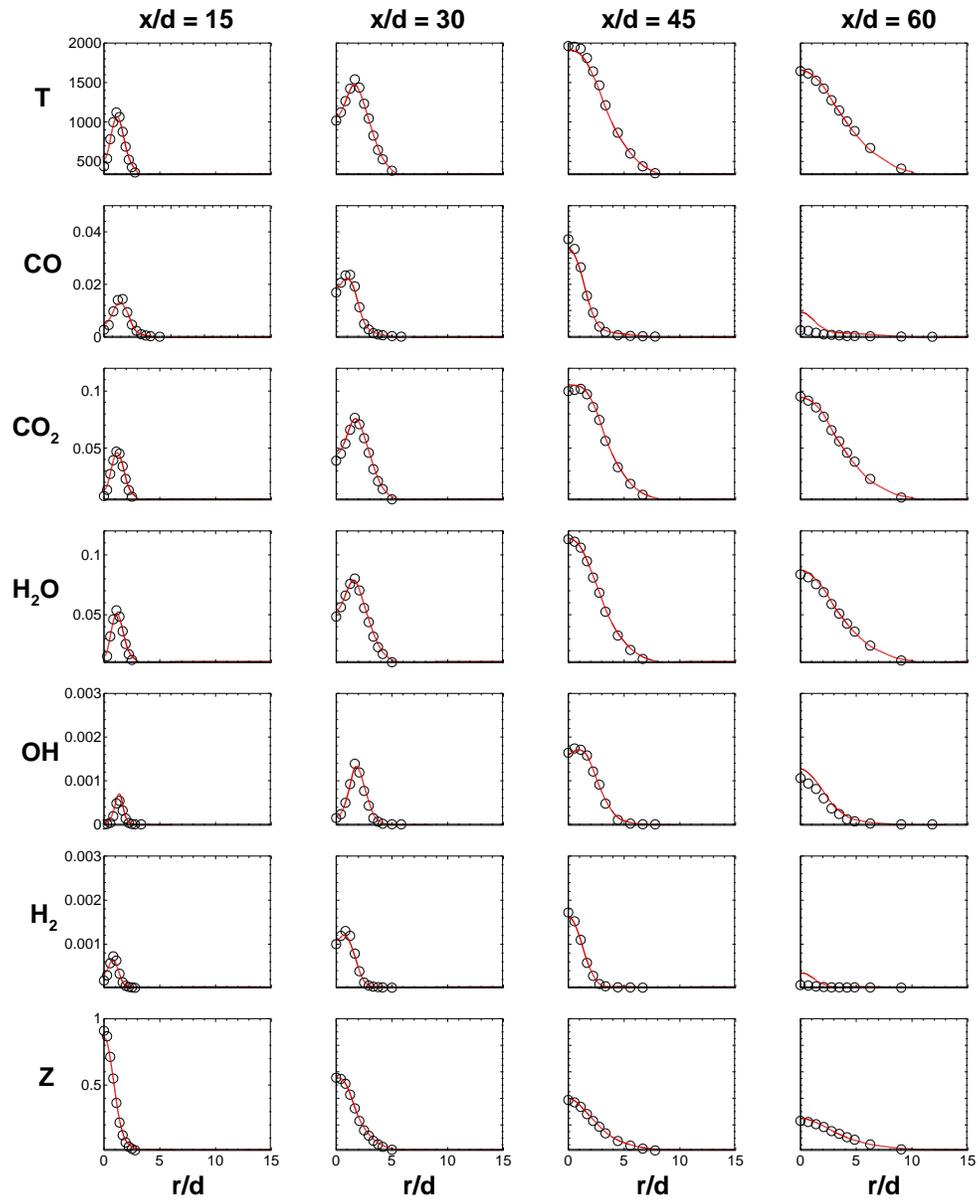

**Figure 11.** Flame F radial profile comparisons of experimental data (symbols) and the closure model (Red solid lines) of the Favre means of temperature, mixture fraction and measured species mass fractions at $x/d$ = 15, 30, 45 and 60.

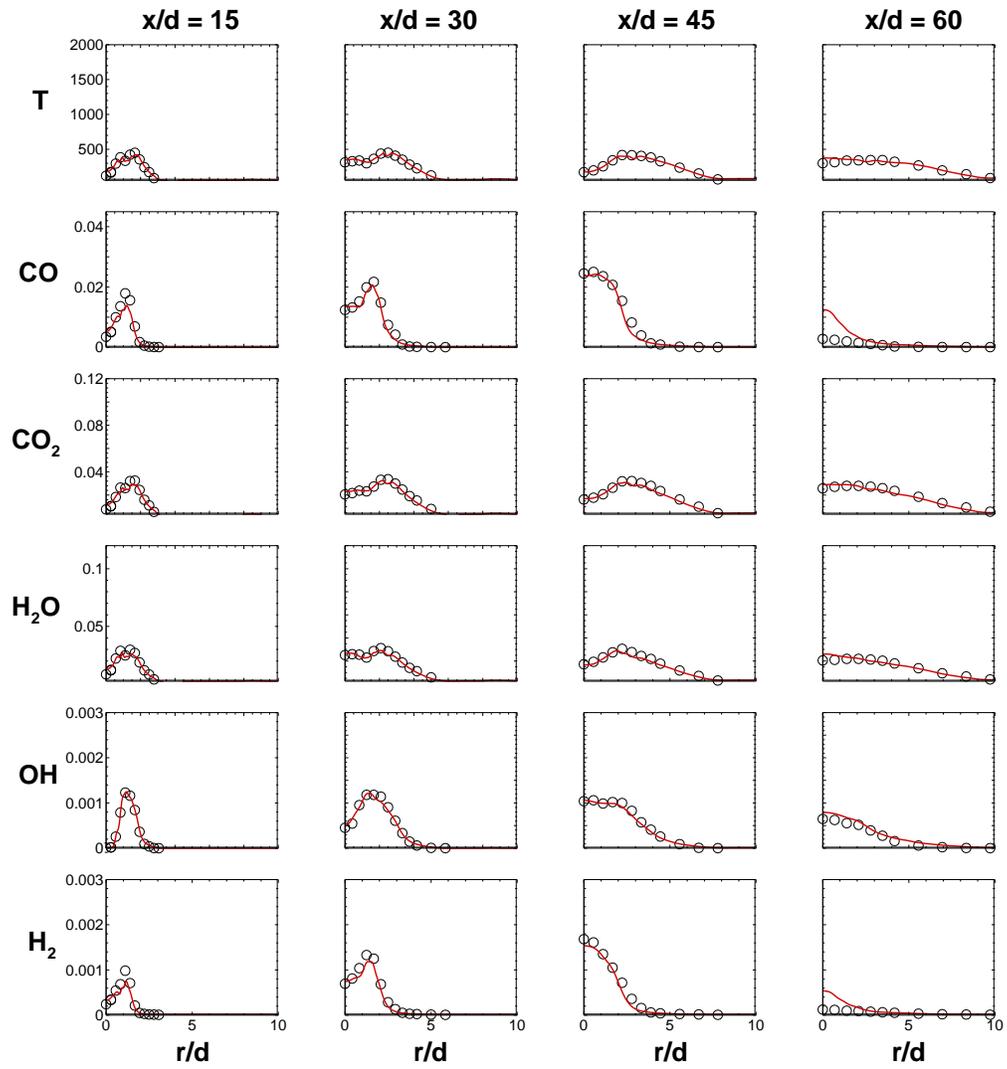

**Figure 12.** Flame D radial profile comparisons of experimental data (symbols) and the closure model (Red solid lines) of the RMS of temperature and measured species mass fractions at $x/d$ = 15, 30, 45 and 60.

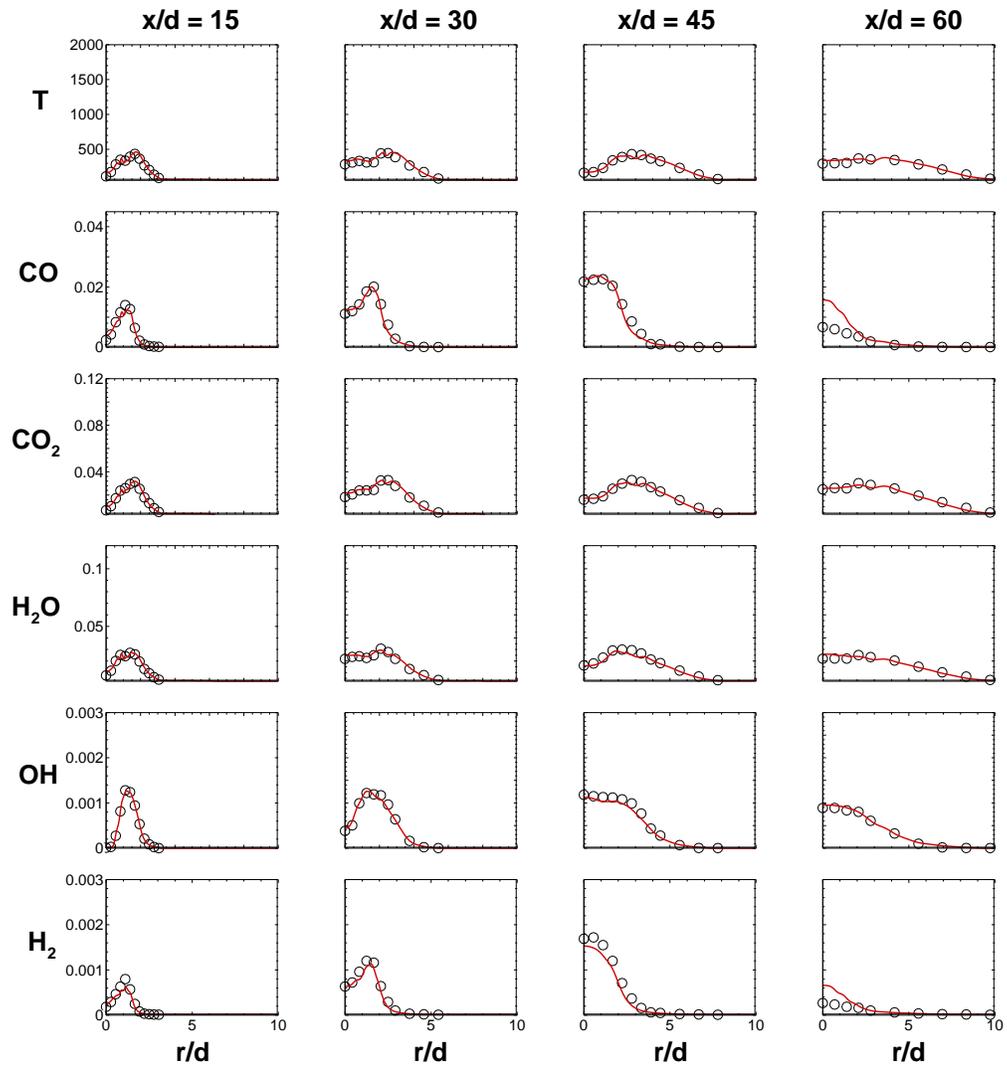

**Figure 13.** Flame E radial profile comparisons of experimental data (symbols) and the closure model (Red solid lines) of the RMS of temperature and measured species mass fractions at $x/d$ = 15, 30, 45 and 60.

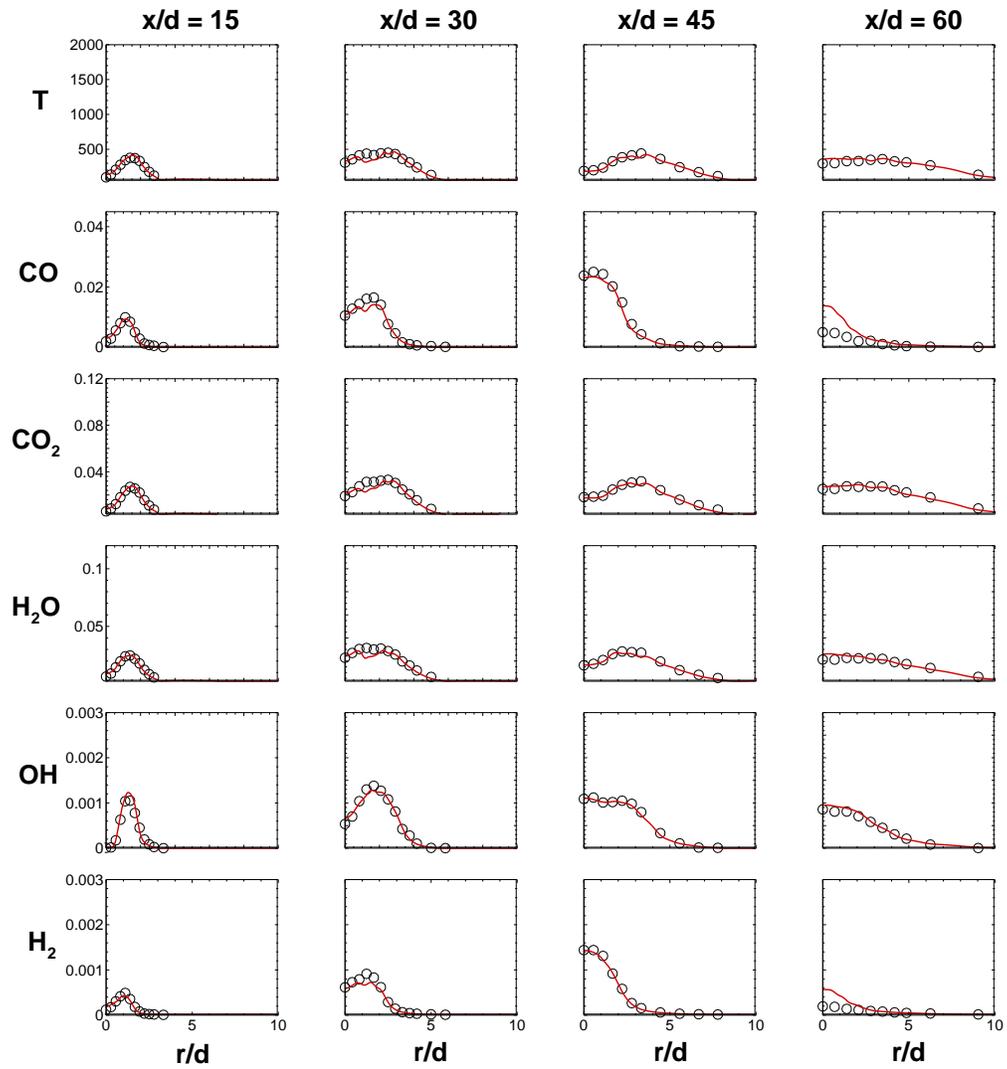

**Figure 14.** Flame F radial profile comparisons of experimental data (symbols) and the closure model (Red solid lines) of the RMS of temperature and measured species mass fractions at $x/d$ = 15, 30, 45 and 60.